\documentclass[prb,aps,twocolumn,showpacs,floatfix]{revtex4}
\usepackage{graphicx}
\usepackage{epsf}
\def\tc{T$_c$}
\def\mgcni3{MgCNi$_3$}
\def\mgcfe3{MgCFe$_3$}
\def\mgcco3{MgCCo$_3$}

\def\prl{Phys. Rev. Lett.}
\def\prb{Phys. Rev. B}
\begin{document}
\title{Study of Normal and Superconducting States of MgCNi$_3$ upon Fe and Co
Substitution and External Pressure}
\author{T. Geetha Kumary, J. Janaki,
Awadhesh Mani, S. Mathi Jaya, V.~\ ~S.~\ ~Sastry, Y. Hariharan, T. S.
Radhakrishnan and M. C. Valsakumar}
\email[email id of the corresponding author: \ ]{geetha@igcar.ernet.in}
\affiliation{Materials Science Division, Indira Gandhi centre for
Atomic Research, Kalpakkam 603 102, India}
\date{\today}
\begin{abstract}
Results of  our study on the superconducting and normal state properties of
the recently discovered superconductor \mgcni3, the effect of Fe and Co
substitution at the Ni site and the effect of pressure are reported. It is
shown that a two band model provides a consistent interpretation of the
temperature dependence of the normal state resistance and the Hall constant.
Whereas band structure calculations suggest an increase in \tc \ upon partial
substitution of Ni with Fe and Co, Co substitution quenches superconductivity
and Fe substitution leads to an increase followed by a decrease in \tc. The
observed variation of \tc \ may be explained in terms of a competition between
increase in \tc \  due to increase in density of states and a decrease due to
spin fluctuations. Based on these results, it is suggested that the spin
fluctuations are weaker in Fe doped samples as compared to the Co doped ones. An
initial decrease in \tc \ (and the normal state resistance) followed by an
increase is observed on application of pressure. The decrease in \tc \ for small
applied pressures can be understood in terms of the decrease in the density of
states at the Fermi level. The subsequent increase in \tc \ with pressure is due
to a lattice softening or a structural phase transition, consistent with the
band structure calculations. It is conjectured that  suppression of spin
fluctuations by pressure may also be responsible for the observed increase in
\tc \ at higher pressures.

\end{abstract}
\maketitle

\section{Introduction}
\label{intro}

The recent discovery\cite{He2001} of  superconductivity in the intermetallic
compound MgCNi$_3$  has initiated a number of studies aimed at
elucidating the nature of its normal and superconducting states. Apart from the
fact that it is the first known superconducting compound which has a perovskite
structure without oxygen, the interest is mainly due to the  preference of
superconductivity over ferromagnetism  which is rather unexpected in a  compound
whose main ingredient is Ni. By proposing an electronic analogy with the
superconducting oxide perovskites, He {\it et. al.}\cite{He2001} suggested that
holes in the Ni d-states might be responsible for electrical conduction in this
compound. The signs of the measured Hall constant\cite{Li2001a} and
thermoelectric power \cite{Li2001b} are, however, consistent with normal state
transport occurring mainly due to electron-like carriers. Most of the normal
state transport properties, like the electrical resistivity $\rho(T)$, Hall
coefficient $R_H(T)$ and thermoelectric power $S(T)$, show unusual temperature
dependence. They are not amenable to a single unified description encompassing
the low and high temperature regimes. For example, the Hall coefficient is
almost temperature independent below 140K, but, its magnitude decreases\cite{Li2001a}
with increase in temperature above 140K. Even though the
conventional Bloch-Gruneisen expression provides a good fit above 70K, a power
law seems\cite{Li2001a} to be better for $\rho(T)$ below 70K. An
electronic crossover occurring at T$_*$ $\sim$ 50K was proposed\cite{Li2001b} to
account for these  anomalies. The normal state NMR properties of MgCNi$_3$ are
also anomalous\cite{Singer2001} and  similar to what is seen in
the exotic superconductor Sr$_2$RuO$_4$. Observation\cite{Mao2001} of a
zero bias conductance peak in the tunneling spectrum is argued to be a signature
of unconventional pairing (like in the case of Sr$_2$RuO$_4$) in this compound.
The NMR measurement\cite{Singer2001} below \tc, on the other hand, shows a
coherence peak whose magnetic field dependence is consistent with
 s-wave pairing. The temperature dependence of the electronic
specific heat below \tc \  is consistent\cite{Lin2002} with the conventional
electron-phonon mechanism  for superconductivity. Analysis\cite{He2001} of the
specific heat data suggests that this compound is a moderately strong-coupled
(electron-phonon coupling constant $\lambda$ $\sim$ 0.77) superconductor. The
observed\cite{Li2001a} temperature dependence of H$_{c2}$ is typical of a
conventional superconductor in the dirty limit. Thus a complete understanding of
the superconducting and normal states of this compound is yet to emerge.

There have been a number of first principle electronic structure
calculations  using Tight Binding Linear Muffin Tin Orbital
(TB-LMTO)\cite{valsa,Dugdale2001,Shim2001}, Full Potential Linear Muffin
Tin Orbital (FP-LMTO)\cite{Shein2001} and Full Potential Linear Augmented Plane
Wave (FP-LAPW)\cite{Singh2001,Kim2002,Rosner2002} methods, within the Local
Density Approximation (LDA), to understand the electronic and magnetic
properties of \mgcni3 \  and related compounds. LDA+U
calculations\cite{Shim2001b} on \mgcni3 \ with U = 5 eV yields essentially the
same band structure as that of the LDA, indicating that electron-correlation
effects of Ni-3d electrons are unimportant in this compound.  This is consistent
with the fact that the measured value of the Wilson ratio\cite{Wilsonratio}  is
only 1.15, typical of a system with modest electron-correlation effects. The
results of these calculations are quite sensitive to the methods employed and
also to the numerical values of the parameters. The general features of
the band structure (which, however, are essentially the same in all these
calculations) can be briefly summarised as follows:  There is a strong
hybridisation between the Ni-3d electrons and the C-2p electrons and thus carbon
plays the crucial role of the mediator of electron hopping. Two of the bands
cross the Fermi level $E_f$ and hence both of them contribute to conduction. The
density of states  at the Fermi level ($N(E_F)$)  is predominantly due to Ni-3d
electrons. One of the distinguishing features of the density of states of this
compound is a sharp peak around 50 meV (its location is very sensitive to the
method of calculation) below $E_F$ arising from the $\pi$ anti-bonding states of
Ni-3d and C-2p states. The Stoner enhancement\cite{Stoner} is moderately
strong to cause ferromagnetic spin fluctuations, but, not strong
enough to obtain a ferromagnetic ground state.

It is well known that the perovskite structure is conducive to a variety of
interesting physical phenomena like superconductivity, magnetism, colossal
magnetoresistance, etc. Soon after the announcement\cite{He2001} of the
discovery of superconductivity in \mgcni3, TB-LMTO calculations were performed
\cite{valsa} to obtain guidelines for synthesis of related compounds with
interesting physical properties. The results of the calculations for the
pristine \mgcni3 \  are consistent with the reports now available in the
literature\cite{Dugdale2001,Shim2001,Shein2001,Singh2001,Kim2002,Rosner2002}.
Our calculations showed that MgCMn$_3$, MgCFe$_3$ and MgCCo$_3$ may form, but
they would be magnetic. Replacement of Mg with other alkaline-earth elements Ca,
Sr and Ba would be interesting. For example, CaCNi$_3$, CaCFe$_3$ and SrCCo$_3$
would be magnetic, whereas SrCNi$_3$, BaCNi$_3$ and CaCCo$_3$ would be
nonmagnetic. As mentioned earlier, band structure calculations showed that the
Fermi level in MgCNi$_3$ lies towards the right side of a peak in the Density of
States (DOS) $N(E)$. Removal of half an electron per formula  unit would shift
the Fermi level to the peak in the DOS. Within the rigid band approximation
(RBA), this can be achieved by substitution of Ni$_3$ by Ni$_{5/2}$Co$_{1/2}$ or
Ni$_{11/4}$Fe$_{1/4}$. Our spin-polarised supercell calculations for
MgCNi$_{3-x}$M$_x$ showed an increase (but much less than the RBA estimate) in
$N(E_F)$, and more importantly, {\it absence} of a magnetic moment, for x=1/4
and 1/2 (M = Co) and x=1/4 (M = Fe). Hence an increase in \tc \  is expected until
x=1/2 for M = Co and x=1/4 for M = Fe. Furthermore, our calculations suggest a
structural instability when the lattice parameter is changed either way. In
particular, a lattice instability is conspicuous  (in the calculation) at a
modest estimated pressure of $\sim$ 2 GPa. Hence effect of application of
pressure in this system would also be interesting. In what follows, we present
the results of our analysis\cite{geethadae,awadheshdae} of the normal state
resistance, the variation in superconducting properties upon Fe and Co
substitution, and the effect of pressure on the superconducting transition
temperature and normal state resistance of MgCNi$_3$.

\section{Sample Preparation and characterization}
\label{prep}

In addition to the pure \mgcni3,  a series of compounds MgCNi$_{3-x}$M$_x$ ( x =
0.05, 0.15 and 0.30 for M = Fe; x=0.05, 0.10, 0.20 and 0.40 for M=Co) were
prepared in order to investigate the effect of hole  doping. The samples were
prepared by following the solid state reaction route described by He {\it et.
al.}\cite{He2001}. Starting materials used were Mg(99.8\%), Ni(99.99\%),
Fe(99.999\%), Co(99.99\%) and amorphous carbon(99\%), all in the form of powder.
20\% excess of Mg is taken in the starting mixture in order to compensate for
the loss of Mg due to evaporation. Also an excess of 50\% C is taken to obtain
the optimum carbon stoichiometry\cite{He2001} in the compound.  The raw
materials were mixed thoroughly and were pressed into pellets. The pellets were
subjected to the heating schedule described by He {\it et. al.}\cite{He2001}, in
an atmosphere of 92\% Ar and 8\% H$_2$.

The crystal structure of all the samples are characterized by x-ray
diffraction. Ac susceptibility and four-probe dc resistance have been measured
in the range 4.2K -300K using home-made dip-sticks. The high pressure
resistance measurement has been  carried out for the pristine MgCNi$_3$ using a
pressure locked opposed Brigman anvils apparatus\cite{awadesh} using the
four-probe method. A superconducting Pb manometer is used for pressure
calibration. We have also characterized the microstructure of MgCNi$_3$ with
SEM.

\section{Results and Discussion}
\label{result}

\begin{figure}
\epsfxsize=7.0cm\epsfysize=5.0cm\centerline{\epsffile{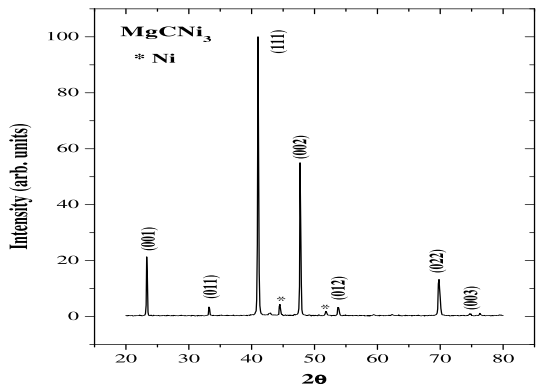}}
\caption{X-ray diffraction pattern of pristine \mgcni3. The lines marked with a *
correspond to unreacted Ni.}
\label{figxray}
\end{figure}

The x-ray patterns of the undoped and doped samples are  consistent with a
perovskite structure with the space group Pm$\bar{3}$m. Small amounts of
unreacted Ni (2-5\%) and MgO impurity ($<$2\%) were present in some of the
samples. Except for these blemishes, all the samples are of good quality, with
sharp x-ray diffraction peaks. The XRD pattern for MgCNi$_3$ is shown in Fig.
\ref{figxray}  as an example. The lattice parameter for the
un-substituted MgCNi$_3$ was found to be 3.810 ($\pm$ 0.0004)$\AA$ which is in
agreement with the values reported in the literature. No significant change in
lattice parameter was noticed upon partial substitution of Ni with Fe or Co.
This observation is consistent with the very small  variation of the
lattice parameter, a(x=0) = 3.806 $\AA$, a(x=3) = 3.802 $\AA$, seen in the
study of MgCNi$_{3-x}$Co$_x$ by Ren {\it et. al.}\cite{Ren2001}.

\begin{figure}
\epsfxsize=7.0cm\centerline{\epsffile{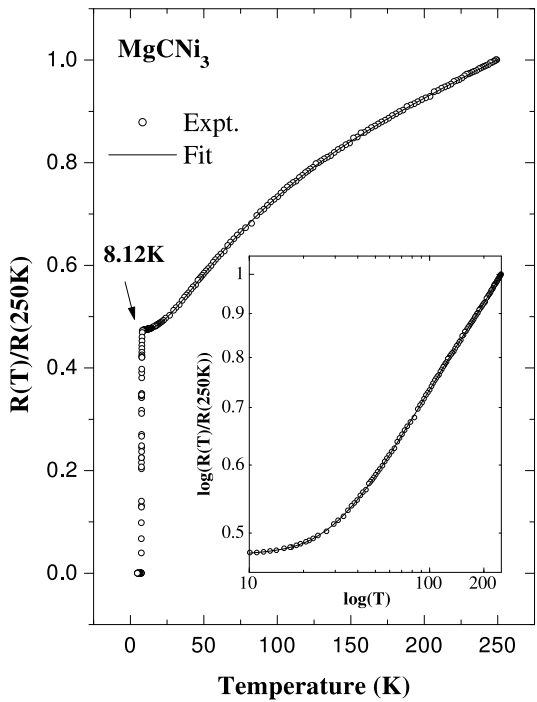}}
\caption{Resistance as a function of Temperature for pristine \mgcni3. The
continuous line is a fit to a two band model of conduction. The normal state
resistance is plotted as a function of temperature on a log-log scale in the
inset to show that the fit is indeed very good in the entire range.}
\label{figrho}
\end{figure}

\begin{figure}
\epsfxsize=7.0cm\centerline{\epsffile{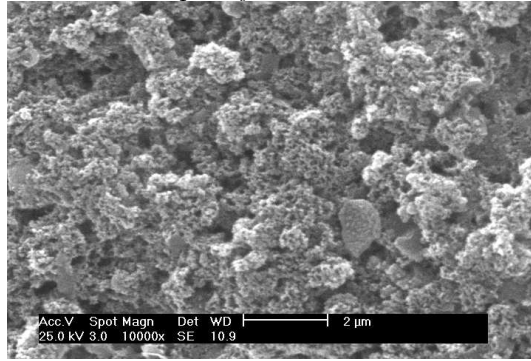}}
\caption{Microstructure of a \mgcni3 \  sample. Fractal like character of the
sample can be seen.}
\label{figsem}
\end{figure}

The variation of resistance R with  temperature T of the pristine   MgCNi$_3$ is
shown in Fig. \ref{figrho}. It can be seen that the {\it shape} of the R(T)
curve presented here is very similar those reported by He et.
al.\cite{He2001} and Li et. al.\cite{Li2001a}. The residual resistivity
ratio of this sample ($R(300K)/R(10K)$) is 2.2 similar to that obtained by He
{\it et. al.}\cite{He2001}.   However, the {\it magnitude} of the
resistivity obtained in this study is approximately 30 times more than that of
He et. al.\cite{He2001} and 10 times that of Li et. al\cite{Li2001a}. This
discrepancy can be understood in terms of the possible differences in the
microstructure that can be expected from the fact that unlike in the study of
He {\it et. al.}\cite{He2001}, our samples were not subjected to high pressure
sintering. Fig. \ref{figsem} shows an SEM pattern of one of the \mgcni3 \
samples. The  fractal character of the sample, which is quite evident from this
photograph, entails a much longer and corrugated path (as compared to the bulk)
for the electrons. This makes the usual procedure of multiplying the resistance
by the (external) Area/Length of the sample to get the resistivity $\rho$
erroneous\cite{badmetal}. Since the actual path length of the electrons in such
a poorly connected medium is longer than that in the bulk, the apparent
resistance will be larger than what is expected from a scaling based merely on
the external geometry\cite{badmetal}. Hence there is a need for a geometric
renormalisation while converting resistance into resistivity in sintered
samples. Our sample is more loosely packed (compared to that of He {\it
et. al.}\cite{He2001}) and hence the larger apparent resistivity of our sample.
A superconducting transition with an onset (mid) \tc \ of 8.13 K (7.69 K) is
observed (90-10\% of the transition width $\Delta$\tc \ = 0.34K). The ac
susceptibility measurement (\ref{figchico}) shows a lower \tc \ (onset \tc \  =
7.35 K) consistent with the values reported in literature \cite{He2001}.

\subsection{Analysis of Normal state Resistance}
\label{resist}

Assuming compliance with Matheissen's rule, we have tried to fit the normal
state resistance of \mgcni3 \  with the expression for a conventional metal,

\begin{widetext}
\begin{equation} R(T) \ = \ A \ + \ BT^2 \
+ \ C \left({{T}\over{\Theta}}\right)^5 \
\int_{0}^{\left({{\Theta}\over{T}}\right)} {dx \ {{x^5}\over{\left[(e^x-1)
(1-e^{-x})\right]}}}
\label{eqbg}
\end{equation}
\end{widetext}
In Eq. \ref{eqbg}, the constant
$A$ represents the  contribution to resistance from electron-impurity
scattering, T$^2$ term accounts for the contribution to resistance from
electron-electron scattering and also from spin fluctuations, and last term is
the Bloch-Gruneisen integral which represents the electron-phonon scattering.
The coefficient $B$ obtained using the fit turns out to be negative and also the
fitted value of the Debye temperature ( $\Theta$ $\sim$ 110K) is too low.
If we substitute this value of  $\Theta$ and the reported value of the
electron phonon coupling constant ($\lambda$ = 0.77) in the McMillan's formula
for \tc \   given by
\begin{equation} T_c \ = \ \left({{\Theta_D}\over{1.45}}\right) \
\exp{\left({{-1.04 (1 + \lambda)}\over{\lambda - \mu^* (1 + 0.62
\lambda)}}\right)},
\label{eqmcmillan}
\end{equation}
we get the superconducting transition temperature \tc \   $\sim$ 4 K, much
less than the experimental value (we have taken $\mu^*$ to be 0.1).

\begin{figure}
\epsfxsize=7.0cm\centerline{\epsffile{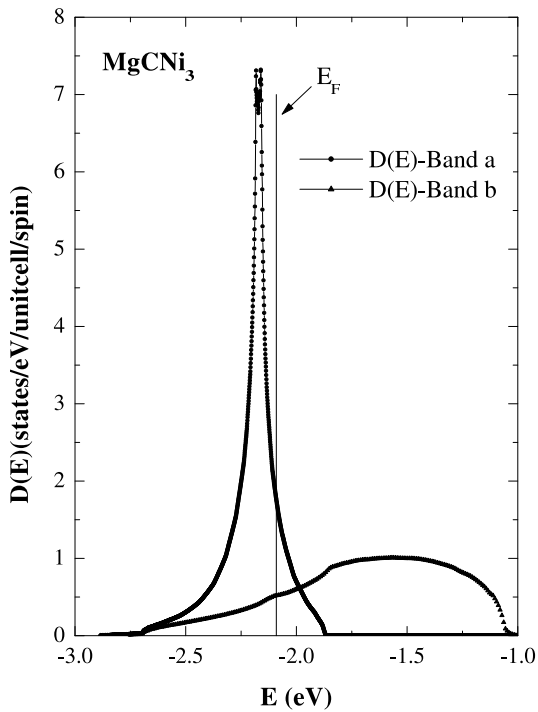}}
\caption{Density of states corresponding to the two bands that cross the Fermi
level. One band is nearly empty and the other is almost full.}
\label{figband}
\end{figure}

This and the reported \cite{Li2001a} temperature variation of the Hall
constant are consistent with a two band model of conduction involving $\lq$light'
electrons coupled to low frequency phonons and $\lq$heavy' holes coupled to high
frequency phonons\cite{valsa}. As mentioned in the Introduction, band
structure calculations show that two bands cross the Fermi level (see
Fig. \ref{figband}) one almost full and the other nearly
empty. There are equal {\it number} of electron-like  and hole-like
charge carriers\cite{numberexplanation} in this compound and the effective mass
of the hole-like carrier is almost ten times that of the electron-like carrier.
Therefore the normal state electrical conduction will be mainly due to electrons
and hence we can expect the Hall constant to be negative. However, both the
electrons and holes will be contributing to the conductivity $\sigma(T)$ and
hence it will be the sum of the conductivities $\sigma_e(T)$ and $\sigma_h(T)$
due to the electrons and holes, respectively. Therefore the effective
resistivity $\rho(T)$ can be written as
\begin{equation} \rho(T) \ = \ A_0 +
{{\rho_e(T) \rho_h(T)}\over{\rho_e(T) + \rho_h(T) )}}
\label{eqtwoband}
\end{equation}
where $\rho_e(T)$ = $1/\sigma_e(T)$ and  $\rho_h(T)$ = $1/\sigma_h(T)$ ,  and
$A_0$ is a constant added to take care of the effect of defects in the
sample. In Eq.\ref{eqtwoband} $\rho_e(T)$ and $\rho_h(T)$ have separate temperature
dependences,
\begin{widetext}
\begin{equation} \rho_{e,h} \ = \ A_{e,h} \ + \ B_{e,h}T^2 \ + \ C
\left({{T}\over{\Theta_{e,h}}}\right)^5 \
\int_{0}^{\left({{\Theta_{e,h}}\over{T}}\right)} {dx \ {{x^5}\over{\left[(e^x-1)
(1-e^{-x})\right]}}}
\end{equation}
\end{widetext}
With $B_e$ and $B_h$ both positive, it is possible to obtain an effective
resistivity such that a fit with Eq. \ref{eqbg} would give  a negative $B$.

The continuous line in Fig. \ref{figrho} is a fit\cite{fitexplanation} of the
normal state resistance, R(T) using the two band model for conduction given by
Eq. \ref{eqtwoband}. It can be seen from the figure that the fit is very good
(better than 0.15\%). $\Theta_e$ obtained from the fit is seen to be smaller
than $\Theta_h$. That is, the electron-like carriers are coupled to low
frequency phonons as compared to hole-like carriers which are coupled to
higher frequency phonons. Assuming the value of the hole-phonon coupling
constant $\lambda$ to be the same as that obtained from the specific heat
measurement\cite{He2001}, we get a \tc\ comparable to the experimental
value. Thus we can infer that superconductivity in \mgcni3 is due to
hole-like carriers. Since $\rho_e(T)$ is much less than $\rho_h(T)$, the
effective resistivity $\rho(T)$ will be close to $\rho_e(T)$ and hence  the
value of $\Theta$ that one gets with a fit of $\rho(T)$ to Eq. \ref{eqbg} will
be close to $\Theta_e$.

The reported variation of the Hall constant $R_H(T)$ can also be
qualitatively explained in the two band model. The Hall constant $R_H(T)$ is the
weighted sum of the Hall constants $R_{He}$ and $R_{Hh}$ due to the  electrons
and holes, respectively,  and it is given by,
\begin{equation}
R_H = {{\left( R_{He} \sigma_e^2 + R_{Hh} \sigma_h^2 \right) }
      \over
     {\left( \sigma_e + \sigma_h\right)^2}}
     = R_{He} {{\left[ 1 - \left( {{\rho_e}\over{\rho_h}} \right) \right]}
     \over
     {\left[ 1 + \left( {{\rho_e}\over{\rho_h}} \right) \right] }},
\end{equation}
Since the number of electrons is equal to the number of holes $R_{Hh}$ =
-$R_{He}$ and hence $R_H$ $<$ 0. If
${{d}\over{dT}}\left({{\rho_e}\over{\rho_h}}\right)$  $>$  0, then the
magnitude of $R_H$ also will decrease with temperature thus explaining the
reported Hall effect behaviour. We thus see that a consistent interpretation of
the temperature dependence of both the normal state resistivity and the Hall
constant can be achieved by invoking a two band model of conduction. It remains
to be seen whether the same physical picture can provide a satisfactory
interpretation of the anomalies seen in the thermopower and NMR relaxation rate.

\begin{figure}
\epsfxsize=7.0cm\epsfysize=5.0cm\centerline{\epsffile{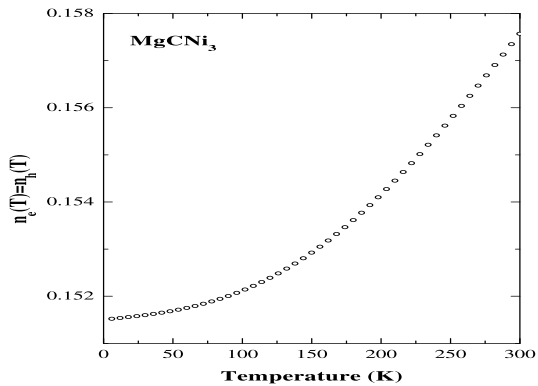}}
\caption{Temperature dependence of the concentration $n_e(T)$ and $n_h(T)$ of
electron-like and hole-like carriers\cite{numberexplanation}, respectively. Both
of these increase by $\sim$ 4\% when temperature increases from 0 K to 300 K.}
\label{figcarrierconc}
\end{figure}

We have also tried to study the effect of the proximity of the peak in the
density of states (van Hove singularity) close to the Fermi level  on the
resistivity. It is seen\cite{valsa} that the chemical potential {\it increases}
with temperature. As a result of this, some of the electrons from the almost
filled band go to the almost empty band ($n_b(T)$ increases at the expense of
$n_a(T)$\cite{numberexplanation}). Thus the numbers of both the electron-like
and hole-like carriers increase with temperature.  As can be seen from Fig.
\ref{figcarrierconc}, carrier concentration increases by approximately 4\% when
the temperature is raised from 0K to 300K. This is one of reasons why the
resistivity is not linear in temperature even in the high temperature regime.

\subsection{Effect of Fe and Co substitution}
\label{subst}

\begin{figure}
\epsfxsize=7.0cm\centerline{\epsffile{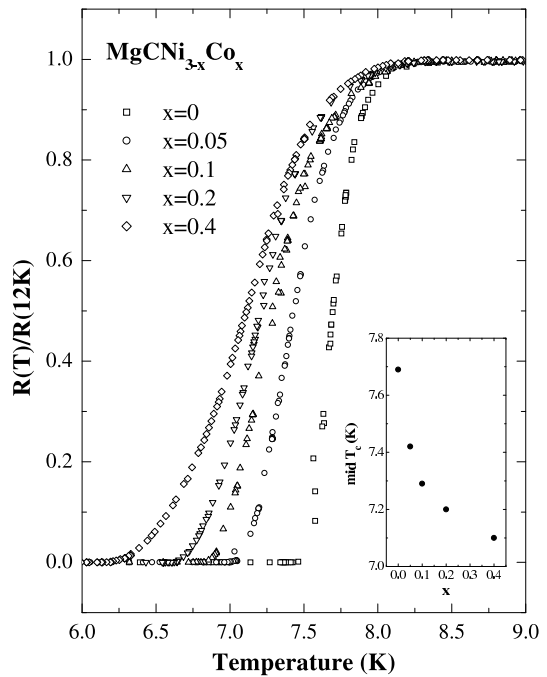}}
\caption{R vs. T for the Co substituted samples for various
concentrations $x$ of Co. \tc \  decreases and the transition width
$\Delta$\tc \  increases monotonically with $x$. \tc \  corresponding to the
midpoint of the resistive transition is shown in the inset.}
\label{figrhoco}
\end{figure}

\begin{figure}
\epsfxsize=7.0cm\centerline{\epsffile{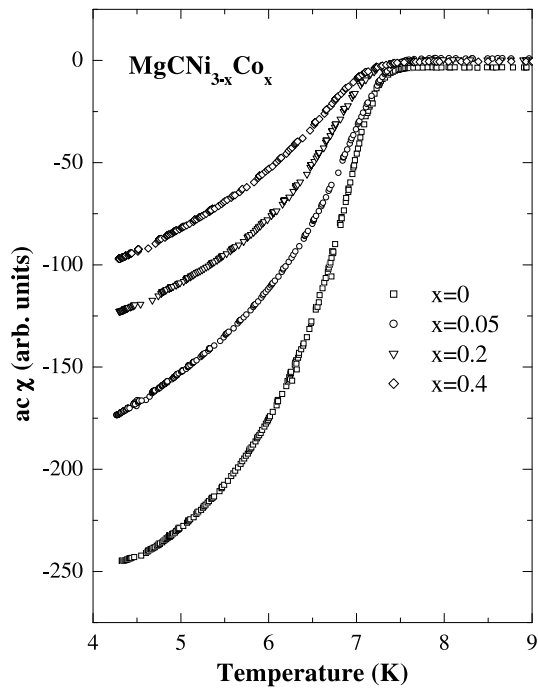}}
\caption{ac $\chi$  vs. T for the Co substituted samples
for various concentrations $x$ of Co. \tc \  as well as the
diamagnetic signal (hence the superconducting volume fraction)
decreases with $x$.}
\label{figchico}
\end{figure}

\begin{figure}
\epsfxsize=7.0cm\centerline{\epsffile{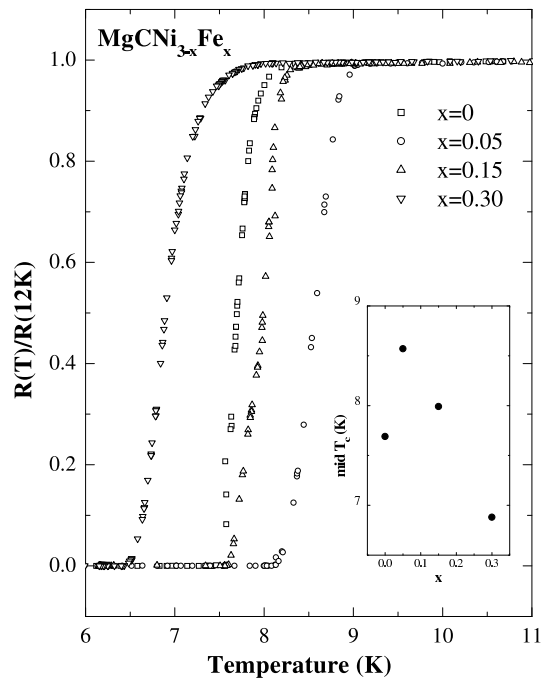}}
\caption{R vs. T for the Fe substituted samples for various
concentrations $x$ of Fe. \tc \  corresponding to the midpoint of the
superconducting transition is shown in the inset. Notice the nonmonotonic
behaviour of \tc}
\label{figrhofe}
\end{figure}

\begin{figure}
\epsfxsize=7.0cm\centerline{\epsffile{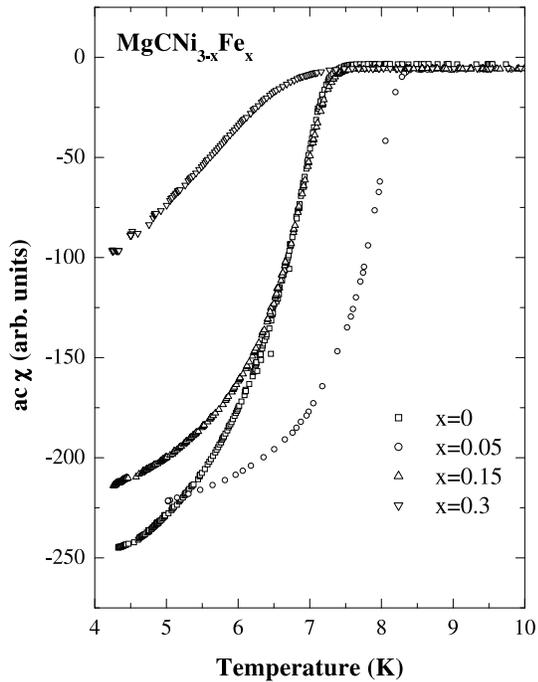}}
\caption{ac $\chi$  vs. T for the Fe substituted samples}
\label{figchife}
\end{figure}

Figs. \ref{figrhoco} and \ref{figchico} represent the superconducting
transitions traced by the resistance and susceptibility measurements
respectively for the Co substituted compounds. The transition temperature
decreases monotonically with increase in Co concentration, similar to the
results now available in literature \cite{Ren2001,Hayward2001}. The variation of
the mid point \tc with respect to x, the Co concentration, is shown as an inset
of Fig. \ref{figrhoco}. The decease in the diamagnetic signal (and hence the
superconducting volume fraction) can also be seen in Fig. \ref{figchico}, which
again is consistent with the results reported in the
literature\cite{Hayward2001}. Fe substitution, on the other hand, leads to an
increase \tc \ followed by a decrease. The superconducting transitions traced
for the Fe substituted compounds are shown in Figs. \ref{figrhofe} and
\ref{figchife}. The maximum \tc \  is obtained when the Fe concentration is
0.05. The onset of the superconducting transition occurs at 9K for this sample
in our resistivity scan. Increasing Fe concentration beyond 0.05 is found to
suppress \tc. The inset of Fig. \ref{figrhofe} represents the mid point \tc \ as
a function of Fe concentration.

An increase in \tc \  was expected for Fe substitution till x=0.25 and for Co
substitution till x=0.5, on the basis of a simple minded interpretation of the
results of the  band structure calculations. However, we observed a monotonic
decrease in \tc \  for Co substitution and an increase followed by a decrease for
Fe substitution. These results are clearly at variance with what is expected
from the band structure calculations. The observed variation in \tc \  can be
explained in terms of an increase in \tc \  due to increase in the density of
states and a decrease due to spin fluctuations: As mentioned in the
Introduction, NMR measurements\cite{Singer2001} have shown existence of
ferromagnetic spin fluctuations in MgCNi$_3$. Also, magnetic susceptibility
measurements have demonstrated that Co doping enhances spin fluctuations in
\mgcni3\cite{Hayward2001}. If we assume the pairing in \mgcni3 is of the
s-wave type, then it will be suppressed by coupling of electrons (more
precisely, the holes) with spin fluctuations. Since we get an increase in  \tc \
for small concentrations of Fe, we believe that the spin fluctuation effect will
be relatively less for Fe substituted compounds as compared to that of Co
substituted ones. This is to be verified experimentally.

\subsection{Effect of Pressure}
\label{pressure}

\begin{figure}
\epsfxsize=7.0cm\epsfysize=5.0cm\centerline{\epsffile{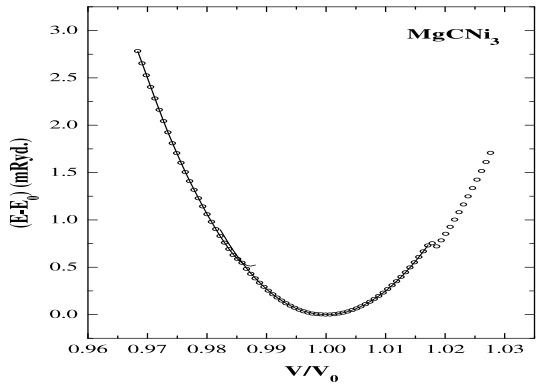}}
\caption{Total energy of \mgcni3 \  as a function of $V/V_0$, where
$V_0$ is the equilibrium volume, obtained from band structure
calculations. Notice the abrupt change of slope near $V/V_0$
$\sim$ .985 as well as the nonmonotonic behaviour near $V/V_0$ $\sim$
1.02}
\label{figtotenergy}
\end{figure}

Figure \ref{figtotenergy} shows the variation of the total energy as a
function of $V/V_0$ ($V_0$ is the equilibrium volume) obtained from our
band structure calculations\cite{valsa}.  The clear change in slope at $V/V_0$
$\sim$ 0.985 is suggestive of a possible lattice instability. The calculated
(using TB-LMTO) bulk moduli of \mgcni3 \  and fcc Ni are 210 and 250 GPa,
respectively. The measured bulk modulus of fcc Ni, on the other hand,  is 180
GPa. Thus the LDA calculation over estimates the bulk modulus. We correct for
this discrepancy by scaling the calculated bulk modulus by a factor 180/250 (the
ratio of the measured and calculated bulk moduli of fcc Ni) and thus estimate
the bulk modulus of \mgcni3 \ to be 151 GPa. With this value of the bulk
modulus, we expect the lattice instability to occur at a pressure $\sim$ 2GPa.
It is known that there is no structural or magnetic phase transitions in \mgcni3
as the temperature is varied, at ambient pressure\cite{Huang2001}. But, there is
no investigation of the structure of \mgcni3 at high pressure. It
would be of interest to do high pressure x-ray diffraction on \mgcni3 to see if
the predicted lattice instability indeed occurs or not.  Here we report the results of
resistivity measurement up to a pressure $\sim$ 7 GPa.

\begin{figure}
\epsfxsize=7.0cm\centerline{\epsffile{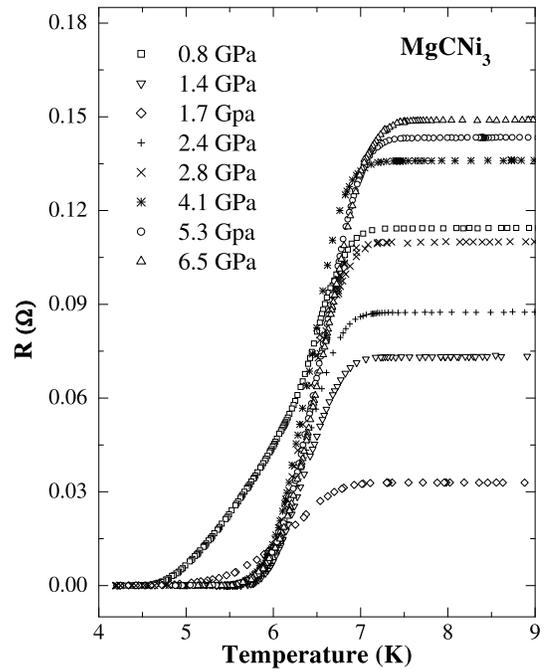}}
\caption{R vs. T of pure \mgcni3 \  for various pressures. The normal state
resistance decreases until a critical pressure $\sim$ 2 GPa and
increases there-after.}
\label{figrhohp}
\end{figure}

\begin{figure}
\epsfxsize=7.0cm\centerline{\epsffile{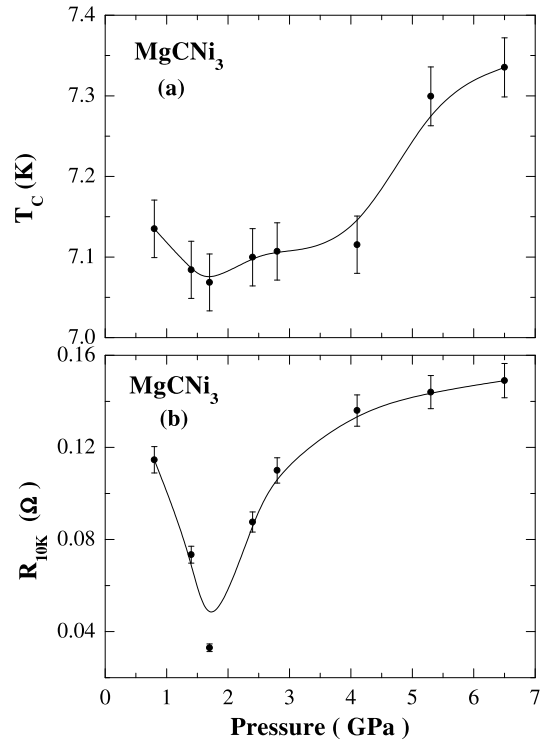}}
\caption{(a) \tc \  as a function of pressure; (b) $R_{10K}$ as a
function of pressure. Notice the correlation between \tc \  and $R_{10K}$.}
\label{figtcp}
\end{figure}

The superconducting transitions and the temperature dependence of
resistivity of \mgcni3 traced for different external pressures are shown in Fig.
\ref{figrhohp}. \tc \  is found to decrease  with increasing pressure
up to $\sim$ 1.7 GPa and it  {\it increases} beyond this pressure. Fig.
\ref{figtcp} (a) depicts the variation of the onset \tc \  with respect to the
applied pressure. The normal state resistance also  shows a similar
trend as can be inferred from Fig. \ref{figtcp} (b) where the variation of the
resistance at 10 K is plotted as a function of the applied pressure. Thus the
variation of \tc \ with pressure is consistent with the behaviour expected of a
superconductor with the conventional electron(hole)-phonon mechanism- larger the
normal state resistivity, higher the \tc.  Band structure calculations  show a
monotonic decrease in $N(E_F)$ with pressure. This would imply a monotonic
decrease in the electron-phonon coupling constant $\lambda$, if there is no
phonon softening.  Fit of the resistance to  Eq.\ref{eqbg} showed an increase
in $\Theta$ with pressure upto 2GPa and a decrease there-after. This implies a
lattice softening commencing at 2GPa. If we assume the usual scaling ($\lambda$
$\sim$ $N(E_F)/\Theta^2$) of the electron(hole)-phonon coupling constant with
the density of states and the Debye temperature $\Theta$, and calculate \tc \
using McMillan's formula, we find a variation of \tc \ with pressure that
qualitatively matches with the experimental observation. It is not yet clear
whether the  decrease in $\theta$ with pressure beyond 2GPa is due to  a
softening of a particular phonon mode or due to a structural phase transition.
Another mechanism that contributes to the increase in \tc \ would be the
suppression of spin fluctuations by pressure, thereby enhancing pairing.

Before concluding, we would like to touch upon the role of carbon, in relation
to superconductivity, in this compound. It is known\cite{Amos2002} that this
compound can form with variable C stoichiometry. However, superconductivity is
absent\cite{He2001,Amos2002} when the C content is less than $\sim$ 0.9. We have
performed band structure calculation (using TB-LMTO) of   MgC$_{0.875}$Ni$_3$
by constructing an appropriate supercell\cite{valsa}. The density of states at
the Fermi level thus calculated is almost the same as that of \mgcni3.
Furthermore, MgC$_{0.875}$Ni$_3$ is non-magnetic as per this spin-polarised
calculation. Thus the absence of superconductivity in \mgcni3 below a C
concentration $\sim$ 0.9 is also surprising. We have, however, found a factor of
two reduction in the height of the peak in DOS close to the Fermi level. This
would support the proposal\cite{Rosner2002}  of a possible connection between
the van Hove singularity and superconductivity in \mgcni3.

\section{Conclusions}

To conclude, we have synthesised \mgcni3 \ and analyzed its  superconducting and
normal state properties. We have also studied the effect of hole doping at
the Ni site and the effect of application of pressure on MgCNi$_3$. It is found
that a two band model for conduction provides a framework for consistently
explaining the observed temperature dependence of the normal state resistance
and the Hall constant. Co substitution quenches superconductivity, whereas Fe
substitution causes an increase in \tc \ followed by a decrease. This may be
understood in terms of a  competition between increase in \tc \  due to
increase in the density of states at the Fermi level and a decrease due to spin
fluctuations. We speculate that spin fluctuations in Fe substituted samples are
weaker than in the Co substituted samples. In fact, it is
known\cite{Hayward2001} that there is an enhancement of spin fluctuations by Co
substitution in \mgcni3. The observed nonmonotonic variation of \tc \ with
pressure may also be pointing out the importance of spin fluctuations vis-a-vis
superconductivity in \mgcni3. The decrease in \tc \ for small applied pressures
can be understood in terms of the decrease in the density of states at the Fermi
level. The subsequent increase in \tc \ with pressure is suggestive of a
lattice softening or a structural phase transition, consistent with the total
energy calculations.  More experimental and theoretical   efforts are needed to
understand these issues better.

\acknowledgements{The authors acknowledge Mrs. M. Radhika, Materials Development
Division, IGCAR for providing the SEM micrographs of the \mgcni3 samples.}


\begin{thebibliography}{99}
\bibitem{Wilsonratio}{Wilson ratio W
=$\left({{\chi}\over{\gamma}}\right)_{expt}/\left({{\chi}\over{\gamma}}\right)_{
band}$ is defined as the ratio the values of
$\left({{\chi}\over{\gamma}}\right)$ obtained from experiment and band theory.
$\chi$ is the susceptibility and $\gamma$ is the coefficient of the linear part
of the electronic specific heat. W =1 for the uncorrelated case and 2 for the
strongly correlated case.}
\bibitem{Stoner}{Stoner enhancement $S$ = $1/(1-\bar{S})$, with $\bar{S}$
= $I_{ex}N(E_F)$, where $I_{ex}$ is the exchange correlation energy.
 The system will have a ferromagnetic instability
whenever $\bar{S}$ $>$ 1.}
\bibitem{numberexplanation}{Band-a in Fig. \ref{figband} is nearly full and
band-b is almost empty. Therefore the number of hole-like carriers $n_h(T)$ =
1-$n_a(T)$ and the number of electron-like carriers $n_e(T)$ = $n_b(T)$,
where $n_a(T)$ and $n_b(T)$ are occupancies (per spin) of band-a and band-b,
respectively. By virtue of the relation $n_a(T)$ + $n_b(T)$ = 1, we find
$n_h(T)$ = $n_e(T)$. }
\bibitem{fitexplanation}{We have fitted the ratio
$R(T)/R(250K)$ so as to eliminate the geometric renormalisation factor that
relates the apparent resistivity to the true resistivity}
\end{thebibliography}
\end{document}